\documentclass[12pt]{article}
\usepackage{graphicx}

\def \b{{\cal B}}
\def \c{{\circ}}
\def \bea{\begin{eqnarray}}
\def \beq{\begin{equation}}
\def \ca{{\cal A}}
\def \eea{\end{eqnarray}}
\def \eeq{\end{equation}}

\def \ol{\overline}
\def \s{\sqrt{2}}
\def \st{\sqrt{3}}

\def \p{\prime}

\def \t{\times}
\def \thet{\theta_\eta}

\begin{document}
\rightline{EFI 10-12}
\rightline{arXiv:1005.2159}
\rightline{July 2010}

\bigskip
\centerline{\bf EFFECT OF $\eta$--$\eta'$ MIXING ON $D \to PV$ DECAYS}
\bigskip

\centerline{Bhubanjyoti Bhattacharya\footnote{bhujyo@uchicago.edu} and
Jonathan L. Rosner\footnote{rosner@hep.uchicago.edu}}
\centerline{\it Enrico Fermi Institute and Department of Physics}
\centerline{\it University of Chicago, 5640 S. Ellis Avenue, Chicago, IL 60637}
\bigskip

\begin{quote}

Charmed meson decays to a light pseudoscalar ($P$) and light vector ($V$) meson
are analyzed taking account of $\eta$--$\eta'$ mixing.  A frequently-used
octet-singlet mixing angle of $19.5^\circ$ is compared with a value of
$11.7^\circ$ favored by a recent analysis of $D \to PP$ decays.
\end{quote}

\leftline{PACS numbers: 13.25.Ft, 11.30.Hv, 14.40.Lb}
\bigskip

Decays of the charmed mesons $D^0$, $D^+$, and $D_s^+$ to a light pseudoscalar
meson $P$ and a light pseudoscalar meson $V$ were analyzed within the framework
of flavor SU(3) in Refs.\ \cite{dpv08} and \cite{Cheng:2010}. A frequently-used
octet-singlet mixing angle between $\eta$ and $\eta'$ of $\thet =
19.5^\circ$ was used in Ref.\ \cite{dpv08}, while Ref.\ \cite{Cheng:2010} used
$\thet = 14.4^\circ$ based on a recent KLOE analysis \cite{KLOE:2009}. In a
study of $D_{(s)} \to PP$ \cite{dpp09}, a best fit to Cabibbo-favored decay
rates was found for $\thet = 11.7^\circ$.  In the present Brief Report
we update fits to $D_{(s)} \to PV$ decays including two decay modes not
considered in \cite{dpv08}, and compare fits based on $\thet = 19.5^\circ$
and $11.7^\circ$.

We review our notation \cite{dpp09}. The angle $\thet$ describing octet-singlet
mixing between $\eta$ and $\eta'$ is defined by
\beq
\eta  = - \eta_8 \cos \theta_\eta - \eta_1 \sin \theta_\eta~,~~
\eta' = - \eta_8 \sin \theta_\eta + \eta_1 \cos \theta_\eta~,~~{\rm where}
\label{eqn:a1}
\eeq
\beq
\eta_8 \equiv (u \bar u + d \bar d - 2 s \bar s)/\sqrt{6}~,~~
\eta_1 \equiv (u \bar u + d \bar d + s \bar s)/\sqrt{3}~,
\label{eqn:a2}
\eeq
Our previous analysis of $PV$ decays utilized $\theta_\eta = \arcsin(1/3) =
19.5^\circ$, for which
\beq
\eta  = (s \bar s - u \bar u - d \bar d)/\sqrt{3}~,~~
\eta' = (2 s \bar s + u \bar u + d \bar d)/\sqrt{6}~.
\label{eqn:b}
\eeq
We consider also $\theta_\eta = 11.7^\circ$, for which an exact fit was found
in Ref.\ \cite{dpp09} to Cabibbo-favored decays.

We refer to Ref.\ \cite{dpv08} for notation.  Amplitudes defined there include
color-favored tree ($T$), color-suppressed tree ($C$), exchange ($E$), and
annihilation ($A$), with a subscript $P$ or $V$ denoting the meson containing
the spectator quark.  Fitting the Cabibbo-favored data
quoted there, we found two solutions (``A'' and ``B''), distinguished by $|T_V|
< |C_P|$ (A) and $|T_V| > |C_P|$ (B).  Fits to singly-Cabibbo-suppressed data
then favored solutions consistent with the set ``A,'' which we shall consider
from now on.  In Table \ref{tab:cfampsA} we show the results of this fit.

We then fit amplitudes involving $\eta$ and $\eta'$, obtaining values for
the amplitudes $T_P$, $C_V$, and $E_V$.  These are compared for the
two-most-favored solutions (denoted by A1 and A2) in Tables \ref{tab:thstd}
and \ref{tab:thnew}.  Predictions for the branching fraction ${\cal B}(D^0
\to \bar K^{*0} \eta')$, listed in the last columns of Tables II and III,
differ slightly between solutions A1 and A2.

\begin{table}[t]
\caption{Solution in Cabibbo-favored charmed meson decays to $PV$ final states
favored by fits \cite{dpv08} to singly-Cabibbo-favored decays.
\label{tab:cfampsA}}
\begin{center}
\begin{tabular}{c c c} \hline \hline
   $PV$   &  Magnitude  &   Relative \\
amplitude & ($10^{-6}$) & strong phase \\ \hline
$T_V$ & $3.95 \pm 0.07$ & --- \\
$C_P$ & $4.88 \pm 0.15$ & $\delta_{C_PT_V} = (-162 \pm 1)^\circ$ \\
$E_P$ & $2.94 \pm 0.09$ & $\delta_{E_PT_V} = (-93  \pm 3)^\circ$ \\
\hline\hline
\end{tabular}
\end{center}
\end{table}

\begin{table}[h]
\caption{Solutions for $T_P$, $C_V$, and $E_V$ amplitudes in Cabibbo-favored
charmed meson decays to $PV$ final states. Solutions A1 and A2 correspond to
$|T_V| < |C_P|$.  Here the $\eta - \eta^\p$ mixing angle is $\thet = 19.5^\c$.
\label{tab:thstd}}
\begin{center}
\begin{tabular}{c c c c c} \hline \hline
   No.  & $PV$  &  Magnitude  &   Relative   & ${\cal{B}}(D^0 \to \ol{K^{*0}}\,\eta\,')$\\
        & ampl. & ($10^{-6}$) & strong phase & ($10^{-4}$) \\ \hline
 A1 & $T_P$ & 7.46$\pm$0.21 & Assumed 0 & \\
    & $C_V$ & 3.46$\pm$0.18 &$\delta_{C_VT_V} = (172 \pm 3)^\c$
      & $1.52 \pm 0.22$\\
    & $E_V$ & 2.37$\pm$0.19 &$\delta_{E_VT_V} = (-110 \pm 4)^\c$ & \\
 A2 & $T_P$ & 6.51$\pm$0.23 & Assumed 0 & \\
    & $C_V$ & 2.47$\pm$0.22 &$\delta_{C_VT_P} = (-174 \pm 4)^\c$
 & $1.96 \pm 0.23$ \\
    & $E_V$ & 3.39$\pm$0.16 &$\delta_{E_VT_V} = (-96 \pm 3)^\c$ & \\
\hline\hline
\end{tabular}
\end{center}
\end{table}

\begin{table}[h]
\caption{Same as Table \ref{tab:thstd} but with $\thet = 11.7^\c$.
\label{tab:thnew}}
\begin{center}
\begin{tabular}{c c c c c} \hline \hline
   No.  & $PV$  &  Magnitude  &   Relative
   & ${\cal{B}}(D^0 \to \ol{K^{*0}}\,\eta\,')$\\
        & ampl. & ($10^{-6}$) & strong phase & ($10^{-4}$) \\ \hline
 A1 & $T_P$ & 7.69$\pm$0.21 & Assumed 0 & \\
    & $C_V$ & 4.05$\pm$0.17 &$\delta_{C_VT_V} = (162 \pm 4)^\c$
      & $1.19 \pm 0.12$\\
    & $E_V$ & 1.11$\pm$0.22 &$\delta_{E_VT_V} = (-130 \pm 10)^\c$ & \\
 A2 & $T_P$ & 5.68$\pm$0.23 & Assumed 0 & \\
    & $C_V$ & 1.74$\pm$0.23 &$\delta_{C_VT_P} = (-162 \pm 6)^\c$
      & $2.19 \pm 0.16$\\
    & $E_V$ & 3.82$\pm$0.15 & $\delta_{C_VT_V} = (-87 \pm 3)^c$ \\
\hline\hline
\end{tabular}
\end{center}
\end{table}

Here we use a line of thought different from the analysis of Ref.\ \cite{dpv08}.
We now calculate global $\chi^2$ values for fits to singly-Cabibbo-suppressed
$D^0 \to PV$ decays for the solutions A1 and A2. We compare the $\chi^2$ values
for the fit with $\thet=19.5^\c$ [including branching fractions $\b(D^0 \to \eta
\omega) = (0.221 \pm 0.023)\%$ \cite{Kass:2009} and $\b(D^0 \to \eta \phi) =
(1.4 \pm 0.5) \times 10^{-4}$ \cite{Amsler:2008} omitted in the original article]
with values for a fit to the same data with $\thet=11.7^\c$. These results are
shown in Tables \ref{tab:oldchi} and \ref{tab:newchi}, respectively.

\begin{table}
\caption{Global $\chi^2$ values for fits to singly-Cabibbo-suppressed $D^0\to
PV$ decays. Also included are the process that contribute the most to a high
$\chi^2$ value.  Here we have taken $\thet=19.5^\c$.
\label{tab:oldchi}}
\begin{center}
\begin{tabular}{l c l c c c} \hline \hline
  No. & Global & \multicolumn{4}{c}{Worst Processes (Highest $\Delta\chi^2$
 values)}\\
      &$\chi^2$& Decay Channel &${\cal{B}}_{th}(\%)$&${\cal{B}}_{expt}(\%)$
 & $\Delta\chi^2$\\ \hline
 A1 & 55.9 &$D^0 \to \eta\,\phi$ &$(4.0\pm0.4)\t10^{-2}$&$(1.4\pm0.5)\t10^{-2}$
 & 16.8 \\
 & & $D^0 \to \eta\,\omega$& $0.33 \pm 0.02$ & $0.221 \pm 0.023$ & 11.3 \\
 A2 & 82.4 &$D^0 \to \eta\,\phi$ &$(5.9\pm0.4)\t10^{-2}$
 &$(1.4\pm0.5)\t10^{-2}$ & 45.8 \\
 & &$D^0 \to \pi^0\,\rho^0$& $0.27 \pm 0.02$ & $0.373 \pm 0.022$ & 10.1 \\
\hline \hline
\end{tabular}
\end{center}
\end{table}

\begin{table}
\caption{Same as Table \ref{tab:oldchi} but with $\thet=11.7^\c$.
\label{tab:newchi}}
\begin{center}
\begin{tabular}{l c l c c c} \hline \hline
  No. & Global & \multicolumn{4}{c}{Worst Processes (Highest $\Delta\chi^2$
 values)}\\
 &$\chi^2$& Decay Channel &${\cal{B}}_{th}(\%)$&${\cal{B}}_{\rm expt}(\%)$
 & $\Delta\chi^2$\\ \hline
 A1 & 35.8 &$D^0 \to \pi^+\,\rho^-$& $0.39 \pm 0.03$ & $0.497 \pm 0.023$
 & 8.5 \\
 & & $D^0 \to \eta\,\omega$ & $0.30 \pm 0.02$ & $0.221 \pm 0.023$ & 6.7 \\
 A2 & 131.4 & $D^0 \to \eta\,\phi$ &$(9.2\pm0.6)\t10^{-2}$
 & $(1.4\pm0.5)\t10^{-2}$ & 100.0 \\
 & & $D^0 \to \pi^0\,\rho^0$& $0.27 \pm 0.02$ & $0.373 \pm 0.022$ & 11.4 \\
\hline \hline
\end{tabular}
\end{center}
\end{table}

We see that solution A1 is favored for both $\thet=19.5^\c$ and $\thet
=11.7^\c$. The solution A2 is disfavored since its prediction for
$\b(D^0 \to \eta \phi)$ is much higher than the experimental value in
both cases. The same conclusion is reached in Ref.\ \cite{Cheng:2010}
for $\thet=14.4^\c$. We will now disregard the A2 solution and only use
the A1 solution for the rest of the analysis.

The next step is to use observed Cabibbo-favored decays to obtain the
annihilation amplitudes $A_P$ and $A_V$ using the amplitudes for
$D_s \to (\bar K^{*0} K^+,\bar K^0 K^{*+}, \pi^+\omega)$ (as quoted in
Table \ref{tab:cfds}) and the A1 solutions. Since we use only 3 independent
inputs to obtain 4 independent parameters (real and imaginary parts of
$A_P$ and $A_V$) instead of obtaining unique solutions, we obtain a
zone of allowed parameter space. We first form a grid of $|A_P|$ and
$|A_V|$ values, and for every point on this grid, use the amplitudes
for $D_s \to (\bar K^{*0} K^+,\bar K^0 K^{*+})$ to obtain the phases
of $A_P$ and $A_V$ relative to $T_V$ (assumed real, as previously.) Thus
for every point on this grid we now have an amplitude for the decay
$D_s \to \pi^+\omega$ using the amplitude representation from Table
\ref{tab:cfds}. We now select only those points on this grid that are
allowed by the experimental value $|\ca(D_s \to \pi^+\omega)|$ including
its one-sigma error bar.

\begin{table}
\caption{Branching ratios \cite{Amsler:2008} and invariant amplitudes for
Cabibbo-favored decays of $D_s$ used to obtain $A_P$ and $A_V$. $\thet$ is
the $\eta - \eta^\p$ mixing angle. $\phi_1 = \arcsin(1/\st) = 35.3^\c$.
\label{tab:cfds}}
\begin{center}
\begin{tabular}{c l c c c c}
\hline \hline
Meson & Decay & Representation
     & ${\cal B}$ \cite{Amsler:2008} & $p^*$ & $|{\cal A}|$ \\
 & mode & & ($\%$) & (MeV) & $(10^{-6})$ \\ \hline \hline
$D_s^+$ & $\ol{K}^{*0} K^+$ & $C_P + A_V$
     & $3.9 \pm 0.6$ & 682.4 & $3.97 \pm 0.31$ \\
 & $\ol{K}^0 K^{*+}$ & $C_V + A_P$
     & $5.3 \pm 1.2$ & 683.2 & $4.61 \pm 0.52$ \\
 & $\pi^+ \omega$ & $\frac{1}{\sqrt{2}}(A_V + A_P)$
     & $0.25 \pm 0.09$ & 821.8 & $0.76 \pm 0.14$ \\
 & $\rho^+ \eta$ & $T_P\cos(\thet+\phi_1) - \frac{A_P + A_V}{\s}\sin(\thet+\phi_1)$
     & $13.0 \pm 2.2$ & 723.8 & $6.63 \pm 0.56$ \\
 & $\rho^+ \eta\,'$ & $T_P\sin(\thet+\phi_1) + \frac{A_P + A_V}{\s}\cos(\thet+\phi_1)$
     & $12.2 \pm 2.0$  & 464.8 & $12.5 \pm 1.0$ \\
\hline \hline
\end{tabular}
\end{center}
\end{table}

\begin{figure}
\begin{center}
\includegraphics[width=1.00\textwidth]{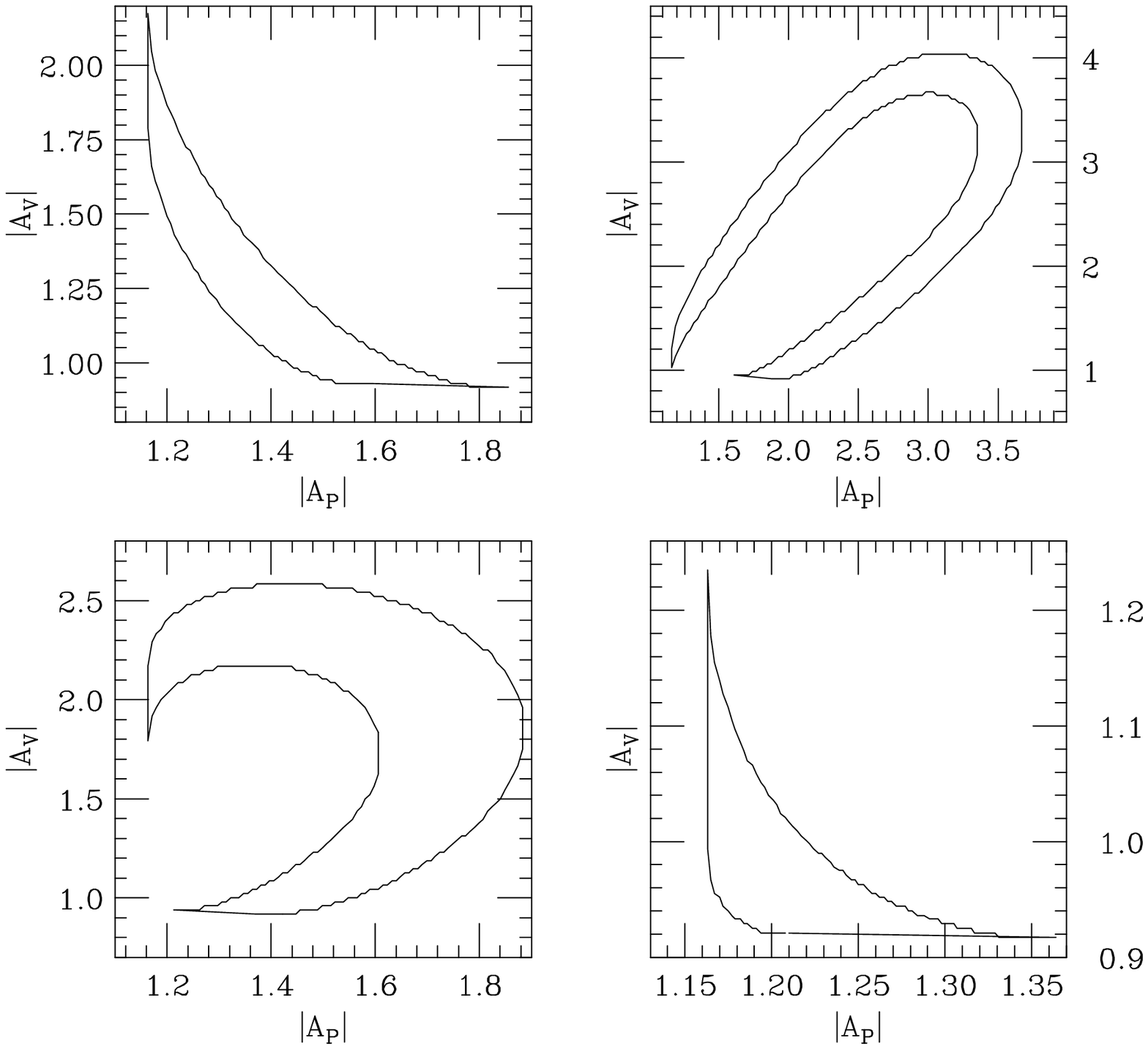}
\end{center}
\caption{Allowed values for $|A_P|$ and $|A_V|$ for $\thet = 19.5^\c$.
In order to obtain the phase of $A_{P(V)}$ we either add (denoted by $+$)
or subtract (denoted by $-$) its phase relative to $C_{V(P)}$ from the phase
of $C_{V(P)}$. Clockwise from top left the 4 panels represent the phase choices:
a) $++$, b) $+-$, c) $--$ and d) $-+$. Thus one may associate a unique value
of the phase of $A_{P(V)}$ with every point on the parameter space in these
plots.
\label{fig:ApAv195}}
\end{figure}

\begin{figure}
\begin{center}
\includegraphics[width=1.00\textwidth]{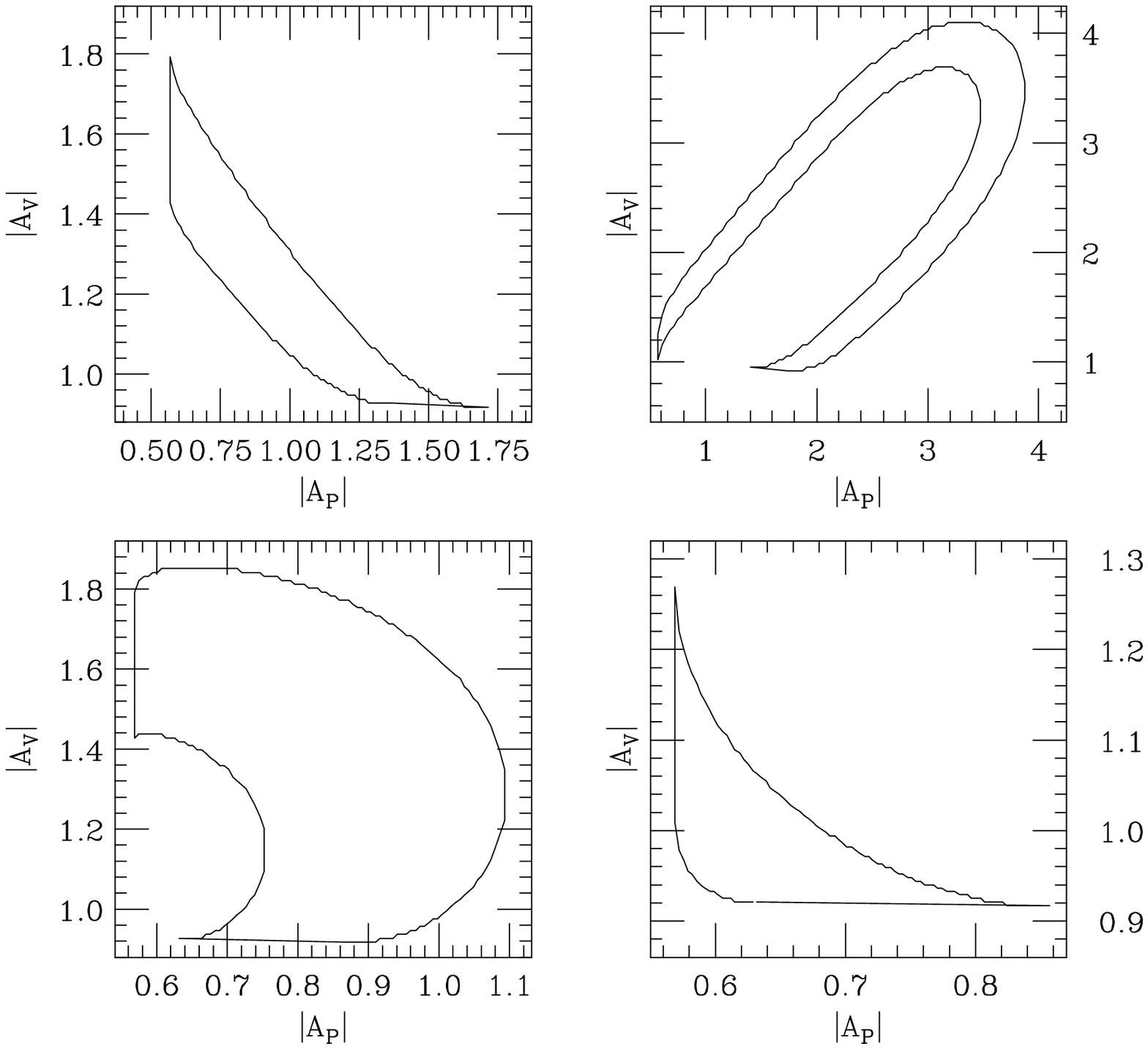}
\end{center}
\caption{Same as Fig. \ref{fig:ApAv195} but with $\thet=11.7^\c$.
\label{fig:ApAv117}}
\end{figure}

Since there is a two-fold discrete ambiguity in choosing the phase of
$A_P$ relative to $C_V$ or that of $A_V$ relative to $C_P$, we obtain
four different sets of allowed zones on the parameter space defined by
$|A_P|$ and $|A_V|$. The allowed zones for $\thet = 19.5^\c$ and $\thet
= 11.7^\c$ are shown in Fig. \ref{fig:ApAv195} and Fig. \ref{fig:ApAv117}
respectively. One may associate unique phases with $A_P$ and $A_V$ (that
may be determined following the method explained above) for every point
on the $|A_P|$ -- $|A_V|$ plane in each of these figures.

To conclude one may now use the range of possible $A_P$ and $A_V$ values
to predict $\b(D_s \to \eta \rho^+)$ and $\b(D_s \to \eta^\p \rho^+)$.
In Ref.\ \cite{dpv08} we used the solution:
\bea
|A_P| &=& 1.36^{+1.16}_{-1.04},~~~~~\delta_{A_P} = (-151^{+83}_{-74})^\c \\
|A_V| &=& 1.25^{+0.34}_{-0.31},~~~~~\delta_{A_V} = ( -19^{+10}_{-9})^\c
\eea
which led us to obtain $\b(D_s \to \eta \rho^+) = (5.6 \pm 1.2)\%$ and
$\b(D_s \to \eta^\p \rho^+) = (2.9 \pm 1.2)\%$. Over the region of allowed
values for $A_{P(V)}$, the central values for these Cabibbo-favored $D_s$
branching ratios vary over the ranges shown in Table \ref{tab:Dsdecay}.

\begin{table}
\caption{Range of predicted branching ratios for $D_s \to (\eta, \eta^\p)\,\rho^+$ using
both $\thet = 19.5^\c$ and $\thet = 11.7^\c$.
\label{tab:Dsdecay}}
\begin{center}
\begin{tabular}{c c c c c}
\hline \hline
Decay & \multicolumn{2}{c}{$\thet = 19.5^\c$} & \multicolumn{2}{c}{$\thet = 11.7^\c$} \\
mode &  Min(\%) & Max(\%) &  Min(\%) & Max(\%) \\ \hline \hline
$\b(D_s \to \eta \rho^+)$ & 3.80 & 6.39 & 6.27 & 8.35 \\
$\b(D_s \to \eta^\p \rho^+)$ & 2.71 & 3.41 & 2.45 & 3.04 \\
\hline \hline
\end{tabular}
\end{center}
\end{table}

The predictions for $\b(D_s \to \eta \rho^+)$ are a bit higher in the new fit using 
$\thet = 11.7^\c$ and slightly closer to the experimental value \cite{Amsler:2008} 
quoted in Table \ref{tab:cfds}. No improvement is seen in the prediction for 
$\b(D_s \to \eta' \rho^+)$ in the new fit using $\thet = 11.7^\c$. The experimental
values for both these branching ratios \cite{Amsler:2008}, as quoted in Table
\ref{tab:cfds}, are much higher than the predictions using this analysis. As mentioned 
in Ref.\ \cite{dpv08}, the relation
\beq
|A(D_s \to \rho^+ \eta')|^2 = |T_P|^2 + |A(D_s \to \pi^+ \omega)|^2
 - |A(D_s \to \rho^+ \eta)|^2
\eeq
is very badly obeyed with the present values of $\b(D_s \to \eta \rho^+)$ and
$\b(D_s \to \eta' \rho^+)$, leading us to suspect either that they have been
overestimated experimentally, or that disconnected diagrams (as studied in
\cite{dpp09}) play a larger role than anticipated. The scarcity of available
data for Cabibbo-favored processes prevents such an analysis in the $D \to PV$
case.

This work was supported in part by the United States Department of Energy under
Grant No.\ DE-FG02-90ER40560.

\end{document}